\documentstyle[epsfig,amssymb]{fbsart11}
\title{Study of the \boldmath $^3{\rm He}({^4{\rm He}},\gamma){^7{\rm Be}}$ 
and \boldmath $^3{\rm H}({^4{\rm He}},\gamma){^7{\rm Li}}$ Reactions in an 
Extended Two-Cluster Model\thanks{Dedicated to Achim Weiguny on the 
occasion of his 65th birthday}  
}
\author{
Attila Cs\'ot\'o\instnr{1},
Karlheinz Langanke\instnr{2}} 
\instlist{
Department of Atomic Physics, E\"otv\"os University, P\'azm\'any 
P\'eter s\'et\'any 1/A, H--1117 Budapest, Hungary \and  
Institute for Physics and Astronomy,  
Aarhus University, DK-8000 Aarhus, Denmark}  
\runningauthor{A. Cs\'ot\'o, K.\ Langanke}
\runningtitle{Study of the $^3$He($^4$He,$\,\gamma$)$^7$Be and 
$^3$H($^4$He,$\,\gamma$)$^7$Li reactions} 

\begin{document}

\maketitle
\begin{abstract}
The $^3{\rm He}({^4{\rm He}},\gamma){^7{\rm Be}}$ and $^3{\rm H}({^4{\rm 
He}},\gamma){^7{\rm Li}}$ reactions are studied in an extended two-cluster 
model which contains $\alpha+h/t$ and $^6{\rm Li}+p/n$ clusterizations. 
We show that the inclusion of the $^6{\rm Li}+p/n$ channels can 
significantly change the zero-energy reaction cross sections, $S(0)$, and 
other properties of the $^7$Be and $^7$Li nuclei, like the quadrupole 
moments $Q$. However, the results agree with the known correlation trend 
between $S(0)$ and $Q$. Moreover, we demonstrate that the value of  
the zero-energy derivatives of the astrophysical S-factors are  more 
uncertain than currently believed. 
\end{abstract}

\section{Introduction}

The $^3{\rm H}({^4{\rm He}},\gamma){^7{\rm Li}}$ and 
$^3{\rm He}({^4{\rm He}},\gamma){^7{\rm Be}}$ reactions play important
roles in astrophysics. The former process is an important ingredient in
big-bang nucleosynthesis, while the latter one is a key reaction in the
solar p-p (proton-proton) chains. The 
$^3{\rm He}({^4{\rm He}},\gamma){^7{\rm Be}}$ reaction produces $^7$Be
in our Sun, which is destroyed by either the capture of an electron or
a proton in the $^7{\rm Be}(p,\gamma){^8{\rm B}}$ process. 
In both cases, neutrinos are generated which are observable by
terrestrial detectors and thus contribute to the solar neutrino problems
\cite{Bahcall}. The production rate of $^7$Be in the Sun is determined
by the competition  of the $^3{\rm He}({^4{\rm He}},\gamma){^7{\rm Be}}$ 
and $^3{\rm He}({^3{\rm He}},2p){^4{\rm He}}$ reactions. Therefore,
the precise knowledge of these cross sections at the most effective
solar energies ($\approx$$\,20$ keV) is an important input to
solar models and to the possible solution of the solar neutrino problems.

There exist several measurements on the $^3{\rm He}({^4{\rm He}},
\gamma){^7{\rm Be}}$ cross section down to energies of about 100 keV. From 
these data, the reaction rate at solar energies is obtained by 
extrapolation. The $^3{\rm He}+{^4{\rm He}}$ fusion reaction at low 
energies is usually viewed as an approximate external capture process 
\cite{Christy}. Thus, the energy dependence of the cross section can then 
be obtained from a simple $^3{\rm He}+{^4{\rm He}}$ potential model, which 
in fact reproduces the energy dependence of the measured data rather well. 
The most recent compilation of the data quotes the $^3{\rm He}({^4{\rm 
He}},\gamma){^7{\rm Be}}$ cross section as $S_{34}(0)=0.53\pm 0.05$ keVb 
\cite{Adelberger}, where the rather small uncertainty mainly reflects 
differences in the absolute normalization of the cross section. As 
customary, we have used the standard S-factor parametrization of the cross 
section (the subscript `34' stands for $^3{\rm He}+{^4{\rm He}}$)
\begin{equation}
S(E)=\sigma (E)E\exp{\Big [2\pi\eta (E)\Big ]}, \hskip 0.5cm
\eta (E)={{\mu Z_1Z_2e^2}\over{k\hbar^2}},
\end{equation}
where $Z_1$ and $Z_2$ are the charge numbers of the fragments in the 
incoming channel, while $\mu$ and $k$ are their reduced mass and wave 
number, respectively.

There exist also more elaborate theoretical studies of the
$^3{\rm H}({^4{\rm He}},\gamma){^7{\rm Li}}$ and 
$^3{\rm He}({^4{\rm He}},\gamma){^7{\rm Be}}$ reactions in the
literature, which are based on the resonating group method or the
related microscopic potential model
\cite{Walliser,Langanke,Mertelmeier,Kajino,Kajinoastro,Altmeyer,Mohr}.
These microscopic studies, based on a single $^3{\rm He}+{^4{\rm He}}$ 
fragmentation, yield an energy dependence of the
$^3{\rm He}({^4{\rm He}},\gamma){^7{\rm Be}}$ cross section, which is 
similar to the potential models. Furthermore as a success of these 
approaches, they correctly predicted \cite{Kajino,Langanke} the low-energy
$^3{\rm H}({^4{\rm He}},\gamma){^7{\rm Li}}$ cross section in absolute
normalization and energy dependence, which has been subsequently
confirmed by the precision measurement of Brune {\it et al.} \cite{Brune}. 

Why do we then want to study these low-energy reactions again? Despite
its apparently successful agreement with data, the dependence of the
cross section on the enlargement of the model space has not been
yet established. A first attempt has been made by Mertelmeier and Hofmann 
\cite{Mertelmeier} who included additionally a $^6{\rm Li}+p$ configuration 
in their model space. This multichannel RGM calculation shows a
somewhat increased S-factor compared to the single-channel case, while the
rise towards lower energies is reduced. Given the
astrophysical relevance it is important to determine whether this weaker
energy dependence is physical or an artifact caused by the approximate
treatment of the asymptotic Whittaker functions in the bound states.
Second, we have found in Ref.\ \cite{B8rad} that taking into account the 
$^6{\rm Li}+p$ clusterization changes some of the key properties of the 
$^7$Be bound states considerably. For example, the $^7$Be quadrupole moment 
is increased by about $0.5-1.0$ $e$fm$^2$ if the $^6{\rm Li}+p$ 
configuration is added to the dominant $^3{\rm He}+{^4{\rm He}}$ 
clusterization.

In the present work we study the $^3{\rm He}({^4{\rm He}},\gamma){^7{
\rm Be}}$ and $^3{\rm H}({^4{\rm He}},\gamma){^7{\rm Li}}$ reactions in an
extended two-cluster model, which can take into account the $^6{\rm
Li}+p/n$ clusterization, in addition to the more important $^4{\rm
He}+{^3{\rm He}}/{^3{\rm H}}$ configurations. Our main goal is to explore 
how much the energy dependence of the low-energy $^3{\rm He}({^4{\rm He}},
\gamma){^7{\rm Be}}$ cross section is affected by the addition of the 
$^6{\rm Li}+p$ configuration. The simultaneous study of the
$^3{\rm H}({^4{\rm He}},\gamma){^7{\rm Li}}$ reaction serves as a
potential check, as in this case the low-energy cross section is 
known rather well. 

\section{Model}

Our study of the $^3{\rm He}({^4{\rm He}},\gamma){^7{\rm Be}}$ and 
$^3{\rm H}({^4{\rm He}},\gamma){^7{\rm Li}}$ reactions is based on the 
microscopic cluster model. As a model space we adopt the dominant 
$^4{\rm He}+{^3{\rm He}}/{^3{\rm H}}$ clusterization, to which the 
$^6{\rm Li}+p/n$ configuration is added as well. For comparison, we also 
perform calculations based solely on the $^4{\rm He}+{^3{\rm He}}/{^3{\rm 
H}}$ clusterization; in this case our studies correspond to the microscopic 
calculations presented in 
\cite{Walliser,Langanke,Mertelmeier,Kajino,Kajinoastro,Altmeyer,Mohr}.
Our wave functions thus have the general form
\begin{equation}
\Psi^{J^\pi}_{^7{\rm Be}}={\cal A}\bigg\{\Big [\Big[\Phi^\alpha\Phi^h
\Big]_{1/2}\chi_L(\mbox{\boldmath $\rho $}_1)\Big ]_{JM}\bigg\}+
\sum_{L,S}{\cal A}\bigg\{\Big [\Big[\Phi^{^6{\rm Li}}\Phi^p\Big]_S
\chi_L(\mbox{\boldmath $\rho $}_2)\Big ]_{JM}\bigg\}
\label{be7}
\end{equation}
and
\begin{equation}
\Psi^{J^\pi}_{^7{\rm Li}}={\cal A}\bigg\{\Big [\Big[\Phi^\alpha\Phi^t
\Big]_{1/2}\chi_L(\mbox{\boldmath $\rho $}_1)\Big ]_{JM}\bigg\}+
\sum_{L,S}{\cal A}\bigg\{\Big [\Big[\Phi^{^6{\rm Li}}\Phi^n\Big]_S
\chi_L(\mbox{\boldmath $\rho $}_2)\Big ]_{JM}\bigg\}.
\label{li7}
\end{equation}
Here ${\cal A}$ is the intercluster antisymmetrizer,
the  internal cluster states $\Phi$ are translationally
invariant harmonic oscillator shell model states ($\alpha
=$~$^4$He, $h =$~$^3$He, and $t =$~$^3$H), the \mbox{\boldmath $\rho $}
vectors are the various intercluster Jacobi coordinates,
$L$ and $S$ are the total orbital angular momentum and spin, respectively, 
$J$ is the total angular momentum, and [...] denotes angular momentum 
coupling. The spin-parity of the $^6$Li ground state is $1^+$, and the 
total parity is $\pi=(-1)^L$. Using (\ref{be7}) or (\ref{li7}) in the
seven-nucleon Schr\"odinger equation, we arrive at an equation
for the intercluster relative motion functions $\chi$. These equations must
be solved with high precision up to large $\rho$ value, because the
electromagnetic transition strengths are shifted towards
large $\rho$ at the
low solar energies. For the $\alpha+h/t$ scattering we neglect the $^6{\rm
Li}$ channels and solve the relative-motion equations by using the
variational Kohn-Hulth\'en method \cite{Kamimura}. For the bound states we
use a version of the so-called Siegert variational method, which was first
applied in cluster model calculations by Fiebig and Weiguny \cite{Fiebig}.
In this method the relative motions $\chi$ are expanded in terms of square
integrable functions (Gaussians in our case) plus a term with the correct
asymptotics. 

We study only E1 transitions here. After setting up the initial scattering 
wave functions and final bound states, the E1 cross section is calculated 
\cite{Mertelmeier,Baye} as 
\begin{equation}
\sigma (E) = \sum_{J_i,J_f} {{1}\over{2}}
{{16\pi}\over{9\hbar}}\left ({{E_\gamma}\over{\hbar c}}\right )^{3}
\sum_{L_\omega}(2L_\omega+1)^{-1}
\vert \langle \Psi^{J_f} \vert \vert {\cal
M}_1^E \vert \vert
\Psi^{J_i}_{L_\omega} \rangle \vert ^2,
\label{sigma}
\end{equation}   
where ${\cal M}_1^E$ is the electric dipole ($E1$) transition operator, 
$\omega$ represents the entrance channel, $E_\gamma$ is the photon energy, 
and $J_f$ and $J_i$ are the total spin of the final and initial state, 
respectively. The initial wave function $\Psi^{J_i}_{L_\omega}$ is a partial 
wave of a unit-flux scattering wave function.

\section{Results}

Meaningful cluster calculations of the $^3{\rm He}({^4{\rm He}},
\gamma){^7{\rm Be}}$ and $^3{\rm H}({^4{\rm He}},\gamma){^7{\rm Li}}$ 
reactions within our chosen model space of $\alpha+h/t$ and $^6{\rm
Li}+p/n$ configurations should satisfy the following conditions: (i)
the experimental value of the $^7{\rm Be}/{^7{\rm Li}}$ ground state energy
with respect to the $\alpha+h/t$ threshold must be reproduced; (ii) the
experimental value of the spin-orbit splitting between the $3/2^-$ and
$1/2^-$ states in $^7{\rm Be}/{^7{\rm Li}}$ has to be correct; (iii) in the
extended model spaces the $^6{\rm Li}+p/n$ threshold must be at the right
position, relative to $\alpha+h/t$. Besides calculations for the
extended model space $\{\alpha+h/t;{^6{\rm Li}}+p/n\}$, we also perform
studies in the single-channel $\alpha+h/t$ model space, corresponding to 
the previous calculations (e.g.\ \cite{Kajino,Altmeyer}). In the later 
case only constraints (i) and (ii) have been fulfilled by adjustments of 
the NN interaction.

The internal structure of the $^6$Li cluster is described 
in our model as an $\alpha+d$ clusterization with an $s$-wave
deuteron and $l=0$ relative motion between the $\alpha$ and $d$
clusters. Varying the size parameters of the $\alpha$ and $d$ clusters 
inside $^6$Li should allow us to satisfy condition (iii), while the 
other two conditions can be met by fine-tuning the effective 
nucleon-nucleon interaction. In our previous multicluster study of the
$^7{\rm Be}(p,\gamma){^8{\rm B}}$ reaction, the Minnesota (MN) 
force \cite{MN} has been identified as the most reliable
interaction. Therefore we will also use it in the present calculation,
even at the expense that it slightly overestimates the total cross
sections for both reactions studied here. In the MN interaction, the
exchange mixture parameter ($u$) and spin-orbit strength are those
parameters that are slightly varied. Additionally, we perform calculations 
with the MHN interaction \cite{MHN} which, as shown by Kajino, reproduces 
the measured cross section data for both reactions in a single-channel 
$\alpha+h/t$ model quite well. 

Choosing the cluster size parameter 
$\beta =0.4$ fm$^{-2}$ and by fine-tuning the MN interaction
we have exactly reproduced the binding energies of the two bound states
in $^7$Be and $^7$Li with respect to the $\alpha$+h/t threshold.
This leads to $u=1.049$ and $V_{\rm SO}=-27.85$ MeV, 
$u=1.044$ and $V_{\rm SO}=-26.7$ MeV, $u=1.008$ and $V_{\rm SO}=-35.94$ MeV,
and $u=0.998$ and $V_{\rm SO}=-36.8$ MeV, respectively for the exchange
mixture parameter of the central force and the strength of the spin-orbit
force in the cases of one-channel $^7$Li, two-channel $^7$Li, one-channel
$^7$Be, and two-channel $^7$Be, respectively. 
For the $^6$Li binding energy we calculate 1.43 MeV, to be compared
with the experimental value of 1.475 MeV \cite{Ajzenberg}. In the case of 
the MHN interaction, by choosing $\beta=0.56$ fm$^{-2}$ and fine-tuning the 
$m$ parameter in the Gaussian of medium-size range (the sum of the Wigner 
and Majorana parameters is kept fixed at the original MHN value), we 
reproduce the binding energies in $^7$Be and $^7$Li, while, however, the 
$^6{\rm Li}+p/n$ threshold energy is overestimated by about 1.5 MeV, 
relative to $\alpha+h/t$; i.e.\ condition (iii) cannot be satisfied. The
corresponding potential parameters, following the same order as above, are 
$m=0.399$ and $V_{\rm SO}=-854.6$ MeV, $m=0.444$ and $V_{\rm SO}=-1393$
MeV, $m=0.399$ and $V_{\rm SO}=-840.4$ MeV, and $m=0.450$ and 
$V_{\rm SO}=-1450$ MeV, respectively, where $V_{\rm SO}$ is the
strength of the long-range spin-orbit force.

\begin{figure}[tb]
\centerline{\psfig{file=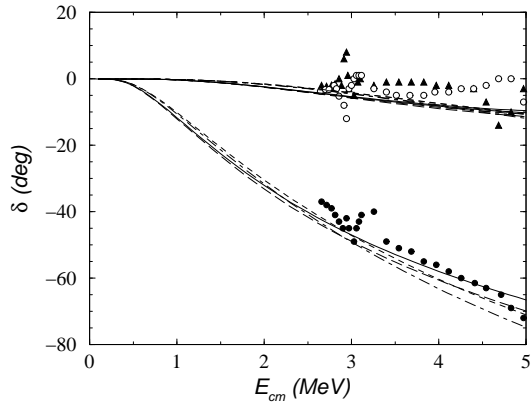,width=7cm}}
\caption{$^4{\rm He}+{^3{\rm He}}$ scattering phase shifts in the 
$J^\pi=1/2^+$ (filled circle), $3/2^+$ (filled triangle), and $5/2^+$ (open
circle) partial waves. The solid (short-dashed) and long-dashed 
(dot-dashed) curves show the results obtained with the MN (MHN) interaction 
for the $\alpha+h$ and $\{\alpha+h;{^6{\rm Li}}+p\}$ model spaces, 
respectively. The calculated phase shifts in the $3/2^+$ and $5/2^+$
partial waves nearly coincide. The data are taken from \cite{Tombrello}.}
\label{fig1}
\end{figure}

In the partial waves which contribute significantly to the capture cross
sections, our interactions give generally more repulsive higher-energy
phase shifts than those indicated by phenomenological analyses. 
As an example, we show our $^4{\rm He}+{^3{\rm He}}$ phase shifts in Fig.\
\ref{fig1}. It is particularly noteworthy that the phase shifts are more 
repulsive in the two-channel calculations than in the one-channel case. This 
is related to the fact that the point-nucleon matter radius of the combined 
nucleus increases by the enlargement of the model space (see Tables 
\ref{tab1} and \ref{tab2} below).

\begin{figure}[tb]
\centerline{\psfig{file=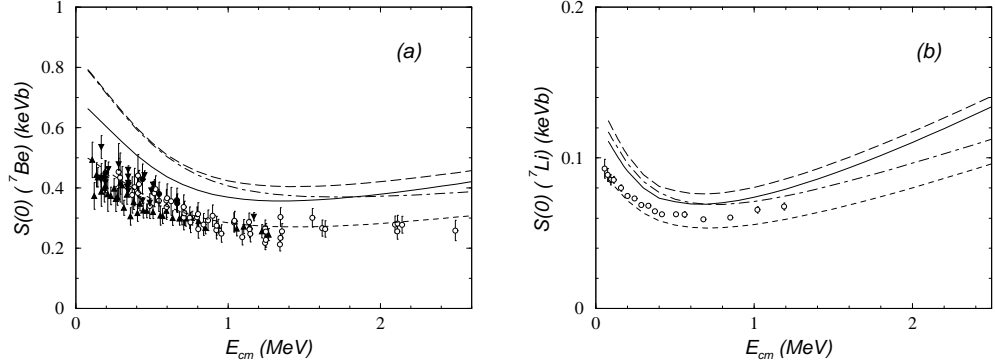,width=13.2cm}}
\caption{Comparison of the calculated astrophysical S-factor $S(E)$ of 
the $^3{\rm He}({^4{\rm He}},\gamma){^7{\rm Be}}$ (a) and 
$^3{\rm H}({^4{\rm He}},\gamma){^7{\rm Li}}$ (b) reactions with data. 
The solid (short-dashed) and long-dashed (dot-dashed) curves show the
results obtained with the MN (MHN) interaction for the $\alpha+h/t$ and 
$\{\alpha+h/t;{^6{\rm Li}}+p/n\}$ model spaces, respectively. The $^7$Be
and $^7$Li experimental data are taken from \cite{Be7exp} and \cite{Brune},
respectively (the symbols follow the notation of Ref.\ 
\protect\cite{Adelberger}).}
\label{fig2}
\end{figure}

The  S-factors, calculated for the extended and single-channel model spaces 
and both interactions, are compared to the data in Fig.\ \ref{fig2}. As 
mentioned above, in the single-channel case the MN interaction overestimates 
the low-energy $^3{\rm He}({^4{\rm He}},\gamma){^7{\rm Be}}$ and 
$^3{\rm H}({^4{\rm He}},\gamma){^7{\rm Li}}$ reactions, while the MHN force 
agrees with the data quite well. However, more importantly for the present
study, the extended model space yields noticeably larger S-factors than in 
the single-channel model (the reason for this increase will be discussed 
below) for both interactions. But we also note that the slope of the
low-energy S-factor changes slightly when comparing the results for the 
two model spaces. This finding can be quantified by the value of
$S^{-1}dS/dE\,(0)$ which decreases from $-0.53$ ($-0.56$) MeV$^{-1}$ to 
$-0.56$ ($-0.63$) MeV$^{-1}$ for the MN (MHN) interaction in the case 
of the $^3{\rm He}+{^4{\rm He}}$ fusion reaction, if we add the $^6{\rm Li}+p$ 
configuration. This result is our first indication that the  reaction 
cannot be treated as a totally external process. As a consequence, the 
usual strategy to adopt the energy dependence of a potential model S-factor 
and normalize it to the measured data becomes slightly questionable as it 
explicitly assumes that the model differs from the data only by the 
asymptotic normalization constant.

In order to explore the origin of the cross section changes between the 
extended and single-channel model spaces, we employ a strategy already 
applied in \cite{Kajino} and, for the $^7{\rm Be}(p,\gamma){^8{\rm B}}$ 
reaction, in \cite{B8rad}. Using the MN interaction for this exploration we 
repeat the calculations for different cluster size parameters ($\beta=0.48$ 
and 0.56 fm$^{-2}$) and try to observe correlations between $S(0)$ and 
other observables in $^7$Be and $^7$Li, respectively. The calculations have 
been performed again in both the single-channel $\alpha+h/t$ and the extended 
$\{\alpha+h/t;{^6{\rm Li}}+p/n\}$ model space. While the above conditions 
(i)$-$(iii) could be fulfilled in each study by a suited adjustment of the 
interaction, the  $^6$Li ground state energy is now underbound relative to 
the $\alpha+d$ threshold; we find $-1.12$ MeV and $-0.39$ MeV for $\beta=0.48$ 
and 0.56 fm$^{-2}$, respectively. As a consequence of this underbinding, the 
$\alpha+d$ separation in the $^6$Li ground state is too large. At the same 
time, the radius of the $\alpha$ particle inside $^6$Li gets smaller, if 
$\beta$ is increased from 0.4 fm$^{-2}$ to 0.56 fm$^{-2}$. These too 
effects together will render the results obtained for $\beta=0.48$ 
fm$^{-2}$ and 0.56 fm$^{-2}$ rather unphysical, but might still allow to 
draw conclusions about the correlations of the cross section with other 
$^7$Be and $^7$Li properties. 

\begin{figure}[!b]
\centerline{\psfig{file=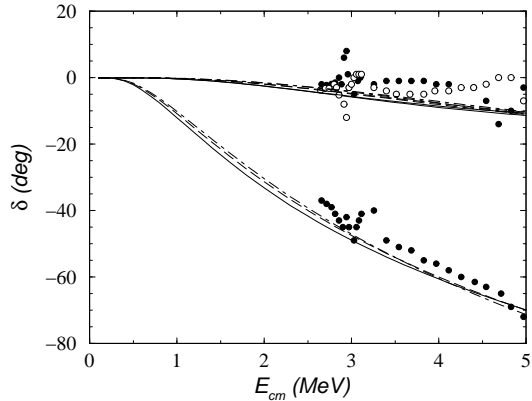,width=7cm}}
\caption{$^4{\rm He}+{^3{\rm He}}$ scattering phase shifts in the 
$J^\pi=1/2^+$ (filled circle), $3/2^+$ (filled triangle), and $5/2^+$ (open
circle) partial waves. The calculations have been performed in the 
$\{\alpha+h;{^6{\rm Li}}+p\}$ model space with the MN interaction and various 
cluster size parameters, $\beta=0.4$ fm$^{-2}$ (solid curves), 
$\beta=0.48$ fm$^{-2}$ (long-dashed curves), and $\beta=0.56$ fm$^{-2}$ 
(dot-dashed curves). The data are taken from \cite{Tombrello}.}
\label{fig3}
\end{figure}

In Fig.\ \ref{fig3} we demonstrate the effect of the cluster size parameter 
on the elastic phase shifts, exemplified for the two-channel study of the
$^3$He+$^4$He reaction. 
We observe that the phase shifts get increasingly more attractive if the 
width parameter $\beta$ increases. This is related to the fact that, with 
increasing $\beta$, the sizes of the combined nuclei $^7$Li and $^7$Be shrink 
(see Tables 1 and 2). Without explicitly showing it in a figure, we note that 
the dependence of the phase shifts on the cluster size parameter is slightly 
stronger in the one-channel case. 

\begin{figure}[tb]
\centerline{\psfig{file=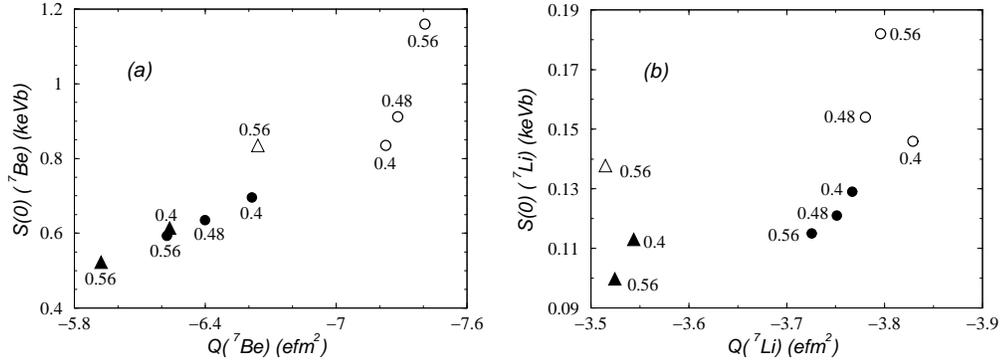,width=13.2cm}}
\caption{The zero-energy astrophysical S-factors of the $^3{\rm 
He}({^4{\rm He}},\gamma){^7{\rm Be}}$ (a) and $^3{\rm H}({^4{\rm He}},
\gamma){^7{\rm Li}}$ (b) reactions as a function of the quadrupole moment 
of $^7$Be and $^7$Li, respectively. The filled and open circles show the 
results for the MN interaction coming from the $\alpha+h/t$ and 
$\{\alpha+h/t;{^6{\rm Li}}+p/n\}$ model spaces, respectively. The triangles
denote the analogous quantities in the case of the MHN force. The values of
the harmonic oscillator size parameter ($\beta$) are indicated.}
\label{fig4}
\end{figure}

Tables \ref{tab1} and \ref{tab2} list our results for the zero-energy cross 
sections, $S(0)$, and the quadrupole moments and point-nucleon matter 
rms radii of $^7$Be and $^7$Li. 
\begin{table}[tb]
\caption{Calculated quadrupole moments ($Q$) and point-nucleon matter rms 
radii
($r$) of $^7$Be, and the zero-energy astrophysical S-factor [$S(0)$] and
its derivative [$S^{-1}dS/dE\,(0)$] of the $^3{\rm He}({^4{\rm He}},
\gamma){^7{\rm Be}}$ reaction in the various model spaces. The entries
denoted by $^*$ are for the MHN interaction, while the rest is for the MN
force.}
\label{tab1}
\begin{tabular}{lr@{}lr@{}lr@{}lr@{}l}
\hline
\tabstrut Model ($\beta$)& \multicolumn{2}{c}{$S(0)$ (keVb)} & 
 \multicolumn{2}{c}{$S^{-1}dS/dE\,(0)$ (MeV$^{-1}$)} &
 \multicolumn{2}{c}{$Q$ 
 ($e$fm$^2$)} & \multicolumn{2}{c}{$r$ (fm)} \\
\hline
$\{\alpha+h;{^6{\rm Li}}+p\}$ & & & & & & & & \\
0.4 & \ \ \ \ \ 0.&83 &\ \ \ \ \ \ \ \ \  $-$0.&57 & \ $-$7.&23 & \ 2.&62  \\
0.48 & 0.&91 & $-$0.&61 & $-$7.&28 & 2.&56 \\
0.56 & 1.&16 & $-$0.&70 & $-$7.&41 & 2.&54 \\
0.56$^*$ & 0.&83 & $-$0.&64 & $-$6.&64 & 2.&43 \\
$\alpha+h$ & & & & & & & & \\
0.4 & 0.&70 & $-$0.&53 & $-$6.&61 & 2.&55  \\
0.48 & 0.&64 & $-$0.&51 & $-$6.&40 & 2.&44  \\
0.56 & 0.&59 & $-$0.&50 & $-$6.&22 & 2.&36 \\
0.56$^*$ & 0.&52 & $-$0.&52 & $-$6.&27 & 2.&51 \\
\hline
\end{tabular}
\end{table}
The $S(0)$ values were determined from fits to the calculated S-factors in
the $E=0.1-0.9$ MeV interval, using the formula given in Ref.\
\cite{Kajinoastro},
\begin{equation}
S(E)=S(0)\exp(-\alpha E)\Big ( 1+a_2E^2+a_3E^3+a_4E^4\Big ).
\label{fit}
\end{equation}
\begin{table}[!b]
\caption{The same as Table \ref{tab1}, except for $^7$Li and the 
$^3{\rm H}({^4{\rm He}},\gamma){^7{\rm Li}}$ reaction.}
\label{tab2}
\begin{tabular}{lr@{}lr@{}lr@{}lr@{}l}
\hline
\tabstrut Model ($\beta$)& \multicolumn{2}{c}{$S(0)$ (keVb)} & 
 \multicolumn{2}{c}{$S^{-1}dS/dE\,(0)$ (MeV$^{-1}$)} &
 \multicolumn{2}{c}{$Q$ ($e$fm$^2$)} & \multicolumn{2}{c}{$r$ (fm)} \\
\hline
$\{\alpha+t;{^6{\rm Li}}+n\}$ & & & & & & & & \\
0.4 & \ \ \ \ \ 0.&148 &\ \ \ \ \ \ \ \ \ $-$2.&26 & \ $-$3.&83 & \ 2.&55 \\
0.48 & 0.&156 & $-$2.&11 & $-$3.&78 & 2.&48 \\
0.56 & 0.&184 & $-$2.&07 & $-$3.&80 & 2.&46 \\
0.56$^*$ & 0.&138 & $-$2.&10 & $-$3.&51 & 2.&36 \\
$\alpha+t$ & & & & & &  \\
0.4 & 0.&131 & $-$2.&19 & $-$3.&77 & 2.&50 \\
0.48 & 0.&122 & $-$1.&98 & $-$3.&75 & 2.&39 \\
0.56 & 0.&117 & $-$1.&82 & $-$3.&73 & 2.&31 \\
0.56$^*$ & 0.&099 & $-$1.&96 & $-$3.&52 & 2.&28 \\
\hline
\end{tabular}
\end{table}
As shown in Fig.\ \ref{fig4}, $S(0)$ and the quadrupole moment $Q$ of 
$^7{\rm Be}/{^7{\rm Li}}$ are nicely correlated in the single-channel  
$\alpha+h/t$ models. 
Although the results of the $\{\alpha+h/t;{^6{\rm Li}}+p/n\}$ model space 
with $\beta=0.4$ fm$^{-2}$ fit well into the trend, 
the results for $\beta=0.48$ and 0.56 fm$^{-2}$  
do not. 
We believe that in the case of $^7$Be this is due to the 
disappearing $\alpha+d$ binding energy inside $^6$Li, while for $^7$Li, it 
is caused  by the adverse behavior of the $\alpha$ and $^6$Li sizes. 

The correlation between $S(0)$ and $Q$, shown in Fig.\ \ref{fig4}, can
easily be understood by recalling that for external capture reactions, at
energies deep below the Coulomb barrier, the $S(0)$ cross section is
practically fully determined by the asymptotic normalization constants of
the bound state wave functions; the $^7{\rm Be}(p,\gamma){^8{\rm B}}$ is a 
well-known example \cite{B8rad}. These normalization constants in turn are 
sensitive mainly to the radii of the effective $\alpha-h/t$ potentials. 
While going from $\beta=0.56$ to 0.4 fm$^{-2}$ in the 
$\alpha+h/t$ model, the sizes 
of the clusters (and thus the radii of the $\alpha-h/t$ potentials) 
increase, which explains the increase in $S(0)$ and $Q$. When the 
$^6{\rm Li}+p/n$ channel is added, $S(0)$ is further increased. Note that
$^6$Li is bigger than $^7{\rm Be}/{^7{\rm Li}}$. Thus the radii of the 
$\alpha-h/t$ potentials are effectively increased by adding the
$^6{\rm Li}+p/n$ component.

As we have already pointed out in Ref.\ \cite{B8rad}, the $^6{\rm Li}+p$
channel has a substantial effect on the quadrupole moment of $^7{\rm Be}$. 
As one can see in Fig.\ \ref{fig4}(a), the inclusion of the $^6{\rm Li}$ 
channel in the $\beta=0.4$ fm$^{-2}$ model results in a roughly 10\% 
increase in $Q$. At the same time, the extension of the $^7{\rm Li}$ 
model space leads to just a 1.6\% increase in the quadrupole moment of 
that nucleus. The explanation of this phenomenon is natural: the 
$^6{\rm Li}+p$ configuration brings in a large charge polarization, which 
strongly affects the quadrupole moment of $^7$Be, while there is no such 
effect for $^7$Li.

In Ref.\ \cite{Kajino} Kajino suggested to determine the zero-energy 
$^3{\rm He}({^4{\rm He}},\gamma){^7{\rm Be}}$ astrophysical S-factor 
indirectly from the correlations between $S_{34}(0)$ and the quadrupole 
moment of $^7$Li. This proposal, however, is spoiled if the $^7$Li 
quadrupole moment is indeed as large as $-4.0$ $e$fm$^2$, as currently 
favored
\cite{Ajzenberg,Pyykko}, and thus noticeably different than the value
of $-(3.4-3.7)$ $e$fm$^2$ \cite{Ajzenberg84}, accepted at the time
Kajino performed his analysis. From Fig.\ \ref{fig4} we observe that 
the large $Q$-value corresponds to a cross section $S(0)\approx 0.19$ 
keVb for the $^3{\rm H}({^4{\rm He}},\gamma){^7{\rm Li}}$ reaction, 
which is about twice the value derived from the precision data of 
\cite{Brune}. This apparent shortcoming might point to the necessity 
to explicitly consider the interference with the internal $^6$Li 
quadrupole moment in cluster studies.

As we mentioned, our $S(0)$ values were determined by fitting the 
calculated $S(E)$ functions by a polynomial form between 0.1 and 0.9 MeV. 
Additionally, Tables \ref{tab1} and \ref{tab2} list the results for 
$S^{-1}dS/dE\,(0)$. In addition to this fit, we tested another functional
form for the low-energy S-factors, suggested in Ref.\ \cite{Jennings}. The 
\begin{equation}
S(E)=n\frac{1+c_1E+c_2E^2}{E+E_B}
\label{pole}
\end{equation}
expression reflects the fact that S has a pole at the  binding 
energy $E_B$ 
of $^7{\rm Be}/{^7{\rm Li}}$, relative to the $\alpha+h/t$ threshold. 
In order to get  good fits even at rather low energies, we
needed to use higher-order terms in the numerator of Eq.\ (\ref{pole}).
This shows the importance of the non-asymptotic contributions to the cross
sections. With a fourth-order polynomial in the numerator in Eq.\
(\ref{pole}), we get practically the same results as shown in Table
\ref{tab1}. One might conclude from our results that the $S_{34}'(0)$ 
values, as determined in Ref.\ \cite{Adelberger} or in Ref.\ \cite{Kajino} 
(for both the $^7$Be and $^7$Li reactions), are more uncertain than 
currently believed. This is in agreement with the recent results of
the NACRE compilation \cite{Angulo}. Here the authors 
recommend to fit the
low-energy $^3{\rm H}({^4{\rm He}},\gamma){^7{\rm Li}}$ S-factor by
the polynomial $S(E)=0.1-0.15E+0.13E^2$. This would imply $S^{-1}dS/dE\,
(0)=-1.5$. This value is quite different from the one ($-2.056$) found in 
\cite{Kajino}. 

\section{Conclusions} 

In summary, we have studied the $^3{\rm H}({^4{\rm He}},\gamma){^7{\rm 
Li}}$ and $^3{\rm He}({^4{\rm He}},\gamma){^7{\rm Be}}$ reactions in an 
extended two-cluster model that can take into account $^6{\rm Li}+p/n$ 
rearrangement channels. We have found that the inclusion of the $^6{\rm 
Li}+p/n$ channel significantly affects the reaction cross sections and 
some key properties of $^7$Be and $^7$Li. For the known correlations 
between the cross sections and, e.g., the quadrupole moments of $^7$Be and 
$^7$Li the results of the extended model (provided the $^6$Li binding 
energy is described sufficiently accurate) fit well into the trend, which 
is predicted by the simpler $\alpha+h/t$ description. We have pointed out 
that the value of $S_{34}(0)$, determined from the $S_{34}(0)-Q^{^7{\rm 
Li}}$ correlation in Ref.\ \cite{Kajino}, seems now to be in conflict with 
the new accepted value of ${^7{\rm Li}}$ quadrupole moment. Finally we have 
discussed that the low-energy slope of the $^3{\rm He}({^4{\rm He}},
\gamma){^7{\rm Be}}$ cross section, i.e. the value of $S_{34}'(0)$, might 
be more uncertain than currently believed.

Further studies in significantly larger model spaces (e.g., using a full
seven-body dynamical description) will be interesting to see how well the
present-day $\alpha+h/t$ cluster models reflect the true nature of the
$^7$Be and $^7$Li nuclei.

\smallskip

\begin{acknowledge}
We thank the Danish Research Council for financial support. The work of 
A.\ C.\ was supported by OTKA Grants F019701 and D32513, FKFP Grant
0242/2000, and by the Bolyai Fellowship of the Hungarian Academy of 
Sciences. 
\end{acknowledge}


\SaveFinalPage
\end{document}